\documentclass[pdflatex,sn-mathphys-num]{sn-jnl}

\usepackage{graphicx}%
\usepackage{multirow}%
\usepackage{amsmath,amssymb,amsfonts}%
\usepackage{amsthm}%
\usepackage{mathrsfs}%
\usepackage[title]{appendix}%
\usepackage{xcolor}%
\usepackage{textcomp}%
\usepackage{manyfoot}%
\usepackage{booktabs}%
\usepackage{algorithm}%
\usepackage{algorithmicx}%
\usepackage{algpseudocode}%
\usepackage{listings}%

\begin{document}

\title[Article Title]{Efficient temporal prediction of compressible flows in irregular domains using Fourier neural operators}

\author[1]{\fnm{Yifan} \sur{Nie}}\email{nyf0002000@gmail.com}

\author*[1]{\fnm{Qiaoxin} \sur{Li}}\email{liqiaoxin@126.com}

\affil[1]{\orgdiv{Department of Mathematics and Physics}, \orgname{North China Electric Power University}, \orgaddress{\street{Beinong Road}, \city{Beijing}, \postcode{102206}, \country{China}}}

\abstract{This paper investigates the temporal evolution of high-speed compressible fluids in irregular flow fields using the Fourier Neural Operator (FNO). We reconstruct the irregular flow field point set into sequential format compatible with FNO input requirements, and then embed temporal bundling technique within a recurrent neural network (RNN) for multi-step prediction. We further employ a composite loss function to balance errors across different physical quantities. Experiments are conducted on three different types of irregular flow fields, including orthogonal and non-orthogonal grid configurations. Then we comprehensively analyze the physical component loss curves, flow field visualizations, and physical profiles. Results demonstrate that our approach  significantly surpasses traditional numerical methods in computational efficiency while achieving high accuracy, with maximum relative $L_2$ errors of (0.78, 0.57, 0.35)\% for ($p$, $T$, $\mathbf{u}$) respectively. This verifies that the method can efficiently and accurately simulate the temporal evolution of high-speed compressible flows in irregular domains.}


\maketitle

\section{Introduction}
\label{sec1}

This study focuses on applying the Fourier Neural Operator (FNO) for modeling the temporal evolution governed by the two-dimensional Euler equations over irregular flow fields. The governing equations are formulated as follows:
\begin{align}
&\frac{\partial\rho}{\partial t}+\nabla\cdot(\rho\mathbf{u})=0\\&
\frac{\partial\rho\mathbf{u}}{\partial t}+\nabla\cdot(\mathbf{u}\otimes(\rho\mathbf{u}))+\nabla p=0\\&
\frac{\partial E}{\partial t}+\nabla\cdot(\mathbf{u}(E+p))=0.
\end{align}
where $\rho$ is the fluid mass density, $\mathbf{u}=(u,v)$ is flow velocity vector, $p$ is the pressure and $E=\rho e+\frac{1}{2}\rho(u^2+v^2)$ is the total energy per unit volume, where $e$ is internal energy.

As fundamental tools for describing physical phenomena, partial differential equations (PDEs) have traditionally been solved using numerical methods such as finite difference, finite element, and finite volume approaches. Despite their maturity, these methods face significant bottlenecks: computational complexity grows drastically with accuracy requirements, mesh generation becomes challenging in complex geometries, and extreme conditions like high-Mach-number flows demand excessively fine grids and tiny time steps, resulting in prohibitive computational costs that hinder real-time simulation and parameter optimization \cite{brenner2008mathematical,weinan2011principles}.

In recent years, deep learning methods have achieved remarkable success in various fields such as computer vision \cite{lecun2015deep} and natural language processing \cite{young2018recent}. These groundbreaking advances have opened up new research directions for the numerical solution of PDEs. The research work in this field can be broadly categorized into three main paradigms: physics-driven methods \cite{zhang2020meshingnet,shen2023physical,ding2023self,xiong2024physics}, physics-constrained methods \cite{wang2021deep,sun2023physics,yang2023physics,du2024physics}, and data-driven methods \cite{xu2021dl-pde,koric2023data,de2023physics,roy2023data}. Among these, data-driven methods possess unique advantages due to their inherent characteristic of learning mapping relationships directly from data without requiring explicit physical priors.

In early data-driven approaches for solving partial differential equations, conventional deep learning architectures have been widely adopted to address computational challenges in physical system modeling \cite{zhang2020physics,yang2020physics,thuerey2020deep,gao2021phygeonet}. These traditional data-driven methods have effectively reduced computational burdens, though the suitability of their inductive biases for PDE solving remains challenging for model interpretability and generalizability.

The field of operator learning has emerged as a transformative approach in scientific machine learning, transitioning from traditional function learning to learning mappings between infinite-dimensional function spaces. The pioneering work of DeepONet \cite{lu2021learning} established one of the first practical frameworks for operator learning, introducing a dual-network architecture that decouples input function processing through branch networks from output evaluation via trunk networks. This innovative structure demonstrated that neural networks could effectively learn nonlinear operators based on the universal approximation theorem for operators. Building upon this foundation, \cite{kovachki2023neural} provided the comprehensive theoretical framework for neural operators, rigorously establishing the mathematical foundations for learning mappings between function spaces with applications to PDEs. Subsequent architectural innovations further advanced the field, with graph neural operators \cite{li2020neural} introducing kernel integration through message passing on graph networks, and multipole graph neural operators \cite{li2020multipole} developing a multi-level framework inspired by classical multipole methods. The FNO \cite{li2021fourier} later emerged as an efficient implementation leveraging spectral methods. This progression demonstrates how operator learning effectively bridges numerical analysis and machine learning, creating new possibilities for scientific computing across diverse physical domains.

The Euler equations, characterized by inviscid assumptions, support for discontinuous solutions, and hyperbolic nature \cite{courant1999supersonic,christodoulou2014compressible}, are valuable for high-speed flow simulations such as in aerospace engineering \cite{pezent2022new}, yet present numerical challenges under high-Mach-number conditions with strong shock waves. Time-dependent PDEs model evolving systems and capture transient phenomena \cite{liu2024multi}, while irregular geometries pose challenges in discretization and boundary treatment \cite{pfaff2020learning}. These combined challenges necessitate innovative approaches for simulating real-world fluid dynamics.

This paper presents a systematic investigation of high-speed compressible flow prediction in irregular domains through input restructuring compatible with FNO requirements. The study encompasses both orthogonal and non-orthogonal mesh configurations. The successful implementation of time bundling techniques has enabled accurate temporal evolution modeling of flow fields. Furthermore, the developed composite loss function has effectively balanced the individual loss components corresponding to different physical quantities, achieving consistent optimization levels across multiple physical parameters.

This work makes the following key technical contributions:

\begin{enumerate}
    \item \textbf{Input restructuring method for irregular flow fields compatible with FNO}: By rearranging flow field data from various mesh types into sequential point arrays, this approach effectively addresses FNO's adaptability challenges on irregular grids, providing a novel data processing paradigm for fluid simulations in complex geometrical configurations.
    
    \item \textbf{Integrated FNO-RNN architecture with temporal bundling}: Incorporating FNO within a recurrent neural network framework and combining it with the temporal bundling technique enables the transition from single-step to multi-step joint prediction, significantly enhancing accuracy in long-term flow sequence simulations and effectively mitigating error accumulation issues.
    
    \item \textbf{Multi-physics balanced composite loss function}: Through systematic analysis of numerical characteristics across different physical quantities, a weighted composite loss function was designed to resolve optimization imbalances caused by magnitude variations, ensuring coordinated accuracy in multi-physics field predictions.
\end{enumerate}

We then conduct a detailed analysis of the physical component loss curves, flow field visualizations, and physical profiles. The results demonstrate that: \textbf{1.} \textbf{efficiency \& accuracy} -- the model's computational efficiency is improved by at least a thousandfold and the error for $\left(p, T, \mathbf{u}\right)$ do not exceed \((0.78, 0.57, 0.35)\%\); \textbf{2.} \textbf{generalization} -- the model can accurately predict the temporal evolution of high-speed compressible inviscid fluids in various irregular flow fields.

The rest of the paper is organized as follows.
In Section~\ref{sec.meth}, we introduced the architecture of FNO, the temporal bundling method, and the composite loss function. Section~\ref{sec.numexp} presents numerical simulations of three distinct irregular flow fields along with comprehensive analysis of the results. Finally, section~\ref{sec.con} concludes the paper by presenting the model's achieved performance and extended applicability, and proposes future research directions.

\section{Methodology}
\label{sec.meth}
\subsection{FNO}
\label{subsec.fno}

The Fourier Neural Operator (FNO) is a neural network architecture designed for learning mappings between infinite-dimensional function spaces. Its core innovation lies in utilizing the Fourier transform to efficiently parameterize the integral kernel in the frequency domain. The architecture primarily consists of three key components: a lifting layer that projects the input function to a higher-dimensional latent space, multiple successive Fourier layers that perform global convolution operations via Fourier transforms, and a projection layer that maps the final representation back to the target output space.

The fundamental building block of FNO is the Fourier layer, which implements the following update rule:
\begin{equation}
	v_{t+1}(x) = \sigma\left(W v_t(x) + \left(\mathcal{K}(\phi) v_t\right)(x)\right)
\end{equation}
where the Fourier integral operator is computed as:
\begin{equation}
\left(\mathcal{K}(\phi) v_t\right)(x) = \mathcal{F}^{-1}\left(R_\phi \cdot \mathcal{F}(v_t)\right)(x)
\end{equation}

In this formulation, $\mathcal{F}$ and $\mathcal{F}^{-1}$ denote the Fourier transform and its inverse respectively, $R_\phi$ represents a learnable transformation in the frequency domain that parameterizes the kernel $\kappa_\phi$, $W$ is a linear weight matrix operating in the spatial domain, and $\sigma$ is a nonlinear activation function. 

In practical implementation, the continuous Fourier transform is discretized and computed using the Fast Fourier Transform (FFT) algorithm. This design enables FNO to capture long-range dependencies efficiently with $\mathcal{O}(n \log n)$ computational complexity. However, this efficient approach necessitates that the input data be defined on uniformly spaced grids to maintain its computational advantages.

To overcome this limitation, we propose a novel approach that rearranges point cloud data from irregular grids into a one-dimensional sequence for FNO processing. As will be demonstrated in subsequent experiments, this methodology exhibits significant advantages in handling complex geometric configurations while maintaining the computational efficiency of the Fourier Neural Operator.

\subsection{Temporal Bundling}
\label{subsec.tb}

To address time series prediction tasks, the conventional approach utilizes the preceding $k$ time steps to predict the current step, i.e., $(t_{n-k}, \dots, t_{n-1}) \rightarrow t_n$. However, this method leads to error accumulation when performing long-term recursive forecasting. To mitigate this issue, we integrate the time bundling method \cite{brandstetter2021message} within a recurrent neural network (RNN) architecture, which enables simultaneous prediction of multiple time steps: $(t_{n-k}, \dots, t_{n-1}) \rightarrow (t_n, \dots, t_{n+k-1})$. This combined approach leverages RNN's temporal modeling capabilities while overcoming error accumulation through multi-step joint prediction. In our experiments, we set $k = 5$.

\subsection{Loss Function}
\label{subsec.lossfuc}

The foundational loss function employed in our approach is the relative $L_2$ loss, defined as:
\begin{equation}
\mathcal{L}_{\text{r}}(\mathbf{y}) =\frac{\|\mathbf{y} - \hat{\mathbf{y}}\|_2}{\|\hat{\mathbf{y}}\|_2}
\end{equation}
where $\mathbf{y}$ denotes the predicted values and $\hat{\mathbf{y}}$ represents the ground truth values.

However, due to the significant numerical disparities among different physical quantities, we apply standardization to the data:
\begin{equation}
\mathcal{J}\left(y_{i,j}\right) = \frac{y_{i,j} - \mu_{i,j}}{\sigma_{i,j}}
\end{equation}
where $\mu_{i,j}$ and $\sigma_{i,j}$ are the mean and standard deviation at $(i,j)$ points computed across the training dataset.

Given that the model needs to predict four physical quantities ($p$, $T$, $u$, $v$), we incorporate weighting coefficients during training to balance their respective contributions. The multi-physics relative $L_2$ loss can thus be formulated as:
\begin{equation}
\mathcal{L}_{\text{multi}}\left(\mathbf{y}\right) = \lambda_p \mathcal{L}_r(p) + \lambda_T \mathcal{L}_r(T) + \lambda_u \mathcal{L}_r(u) + \lambda_v \mathcal{L}_r(v)
\end{equation}
where $\lambda_p$, $\lambda_T$, $\lambda_u$, and $\lambda_v$ are tunable hyperparameters that control the relative importance of each physical quantity in the overall optimization objective.

The composite loss function is constructed by integrating additional regularization terms and physical constraints:
\begin{equation}
\text{Loss} = \alpha\mathcal{L}_{\text{multi}}(\mathbf{y}) + \beta\mathcal{L}_{\text{multi}}\left(\mathcal{J}\left(\mathbf{y}\right)\right)
\end{equation}
where $\alpha$ and $\beta$ are balancing coefficients that control the relative importance of each component in the overall optimization objective.

\section{Numerical Experiments}
\label{sec.numexp}
In this section, experiments were conducted on three irregular domains, covering both orthogonal and non-orthogonal mesh configurations.
\subsection{Forward Step}
\label{subsec.fwdstp}
\begin{figure}[!ht]
\centering
\includegraphics[width=0.5\textwidth]{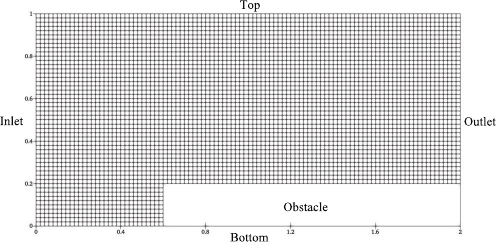}
\caption{Computational grid and boundary labels of Forward step}
\label{fig.fwdstp}	
\end{figure}

The configuration of the flow field is illustrated in fig.\ref{fig.fwdstp}. It is formed by subtracting a smaller rectangular section measuring $1.4 m$ in length and $0.2 m$ in height from a base rectangle with dimensions of $2 m \times 1 m$. Within this defined geometry, the flow conditions are specified as follows: gas enters uniformly from the left boundary, serving as the inlet, and exits through the right boundary, designated as the outlet. The upper and lower boundaries are both configured as solid walls.

The computational domain is discretized using a structured grid with 100 cells in the $x$-direction and 50 cells in the $y$-direction. The fluid is modeled as a normalized inviscid gas with a speed of sound of 1 m/s at 1 K and a specific heat ratio $\gamma = 1.4$. This gas model is employed consistently throughout the subsequent discussion. The initial conditions (internal field) and boundary conditions are detailed in Table~\ref{table.fwdstpconfiguration}. Both the internal field and inlet are initialized with a velocity component $u = 2.0-4.0$ m/s, $v = 0$ m/s, pressure $p = 0.7-1.3$ Pa, and temperature $T = 0.7-1.3$ K, while the boundary conditions are configured as follows: the outlet employs \textit{zeroGradient} for both velocity and temperature, and \textit{waveTransmissive} for pressure; the top and bottom use \textit{symmetryPlane}; the obstacle applies \textit{slip} condition for velocity with \textit{zeroGradient} for pressure and temperature.

\begin{table}[!ht]
    \centering
    \small
    \caption{Boundary conditions and initial field configuration of Forward step. The abbreviations WT, ZG and SP represent waveTransmissive, zeroGradient and symmetryPlane boundary conditions respectively}
    \begin{tabular}{lcccc}
    \toprule
        Boundary & $p$ (Pa) & $T$ (K) & $u$ (m/s) & $v$ (m/s) \\ 
    \midrule
        Internal Field/Inlet & 0.7--1.3 & 0.7--1.3 & 2.0--4.0 & 0  \\ 
        Outlet & WT & ZG & ZG & ZG \\ 
        Obstacle & ZG & ZG & slip & slip \\ 
        Top/Bottom & SP & SP & SP & SP \\ 
    \bottomrule
    \end{tabular}
    \label{table.fwdstpconfiguration}
\end{table}

The flow field data is generated using the fluid solver in \textit{OpenFOAM} v11. The computational time step is set to $2 \times 10^{-3}$~s, with the simulation running from 0 to 2~s. Numerical results are sampled every 0.2~s, thereby generating a dataset composed of approximate solutions to the Euler equations. The dataset is partitioned into a training set (4050 samples), a validation set (450 samples), and a test set (700 samples).

\begin{figure}[!ht]
\centering
\includegraphics[width=0.5\textwidth]{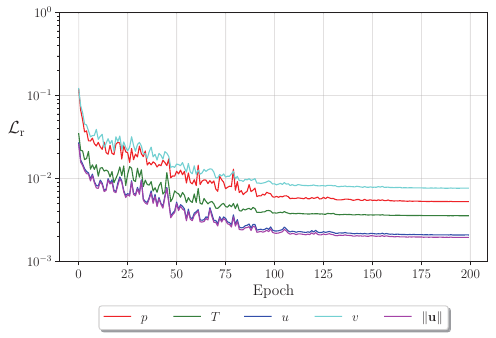}
\caption{Errors of Forward step}
\label{fig.fwdstp_loss}	
\end{figure}

In the initial experiments, due to the simplicity of the flow field, we did not modify the hyperparameters of the loss function, setting $\lambda_p = \lambda_T = \lambda_u = \lambda_v = 1$, $\alpha = 1$, and $\beta = 0$. As shown in Fig.~\ref{fig.fwdstp_loss}, $\mathcal{L}_{\text{r}}$ of all physical components remain within acceptable ranges. Since $v$ constitutes a relatively small component of $\mathbf{u}$, the $\mathcal{L}_{\text{r}}$ values for $u$ and $\|\mathbf{u}\|$ are very close.

\begin{figure}[!ht]
\centering
\includegraphics[width=0.5\textwidth]{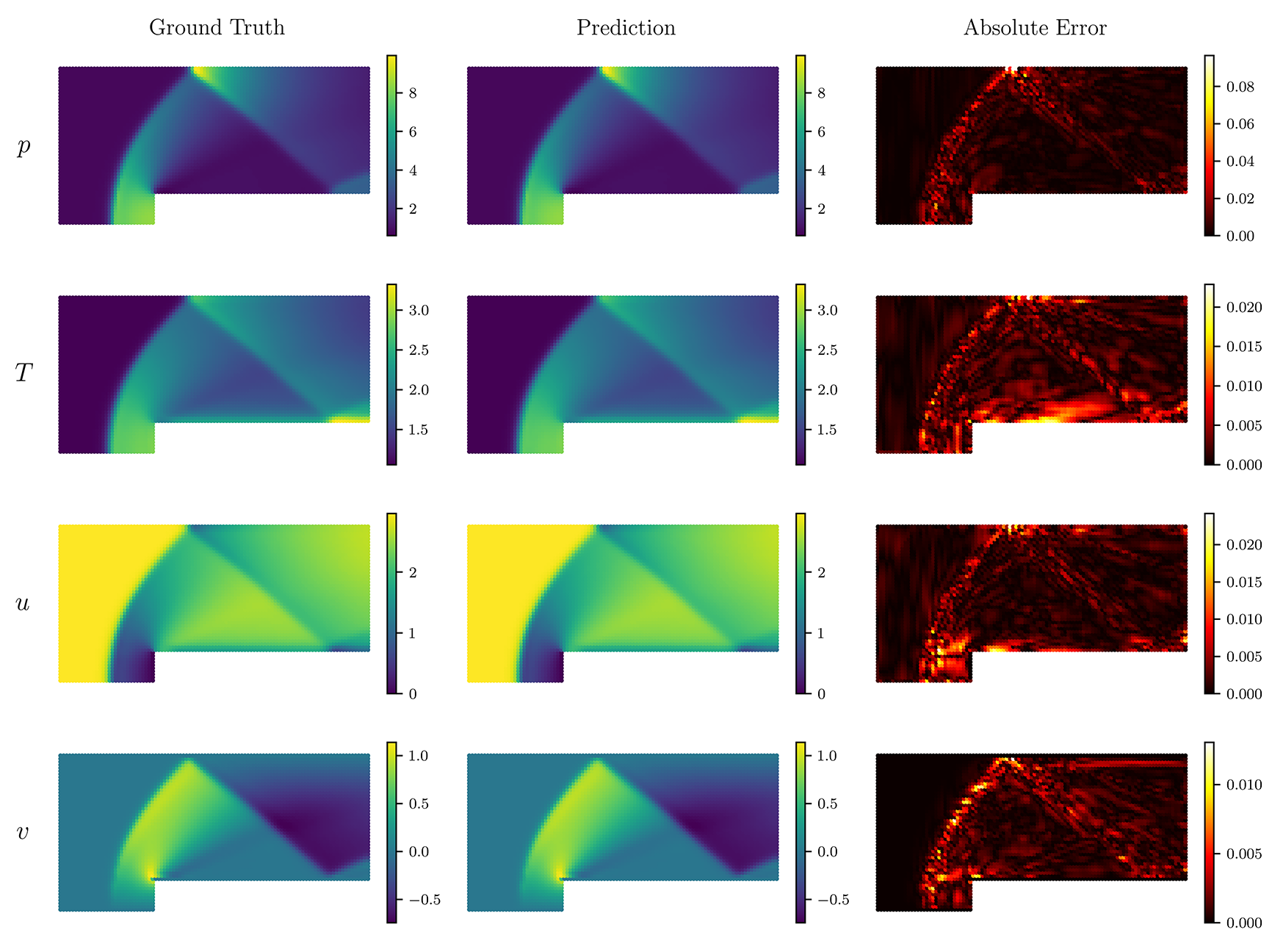}
\caption{Physical field and absolute error of Forward step at $t = 1.8$ s}
\label{fig.fwdstp_t18}
\end{figure}

Figure~\ref{fig.fwdstp_t18} displays the ground truth, prediction, and absolute errors of the entire physical fields ($p$, $T$, $u$, $v$) at time $t = 1.8$ s. It can be observed from the figure that the predictions are close to the ground truth, with errors primarily concentrated at the shock locations. The maximum relative error at the most discrepant regions does not exceed 2\%.

\begin{figure}[!ht]
\centering
\includegraphics[width=0.5\textwidth]{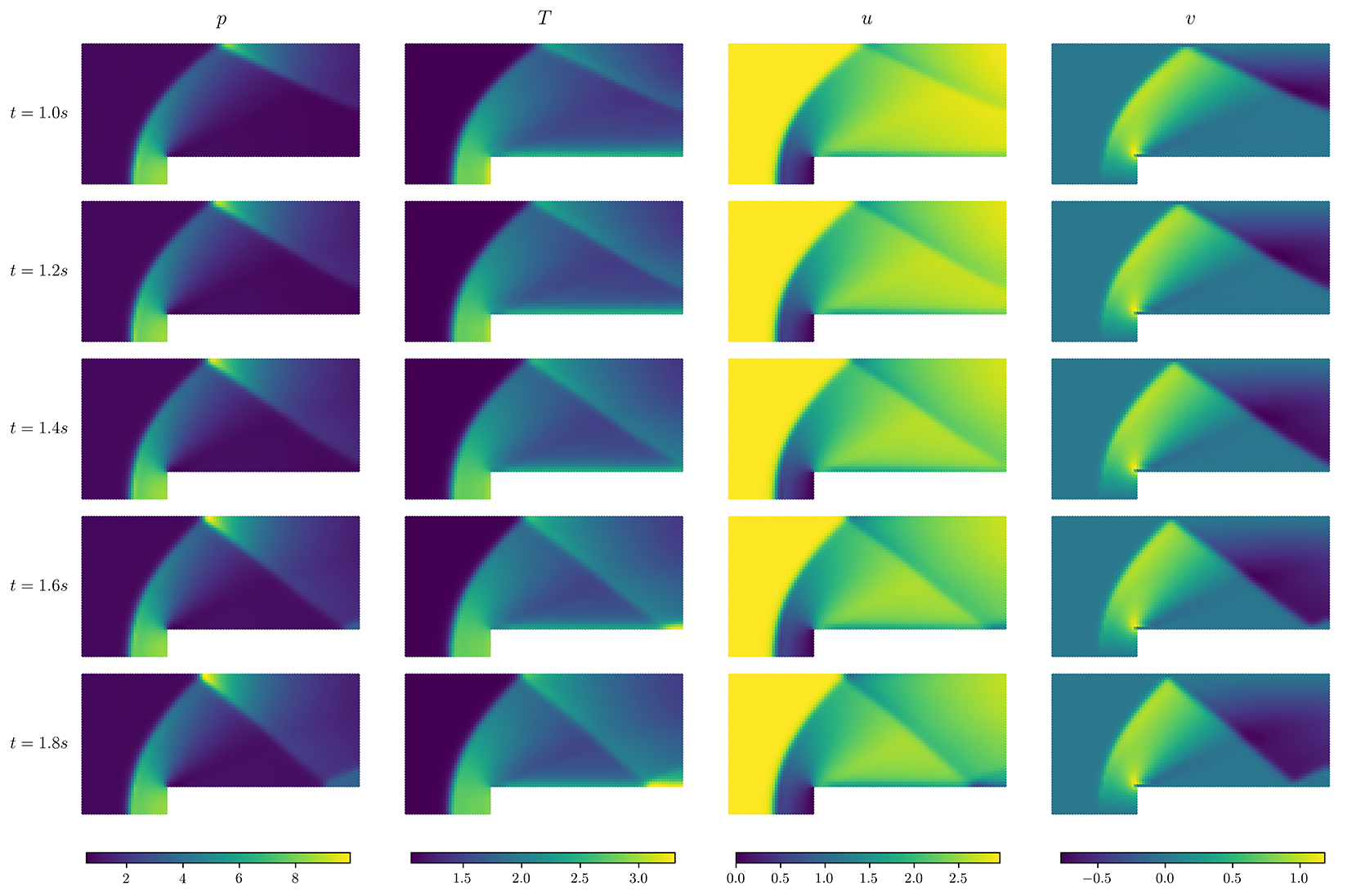}
\caption{Multi-physics field predictions across $t=1.0-1.8$ s of Forward step}
\label{fig.timeline_of_fwdstp}	
\end{figure}

Figure~\ref{fig.timeline_of_fwdstp} presents a visualization of the flow field predicted by the model. This figure provides a comprehensive visual analysis of the temporal evolution of pressure, temperature, and velocity components, arranged in five rows and four columns. Notably, the obstruction created by the obstacle at the bottom of the flow field causes gas compression and shock wave formation on its left side. Subsequently, this shock wave reflects from the top boundary and begins to undergo secondary reflection from the upper surface of the obstacle starting at $t = 1.6$ s. These results demonstrate that the model accurately captures complex phenomena such as shock wave formation and reflection.

\begin{figure}[!ht]
\centering
\includegraphics[width=0.5\textwidth]{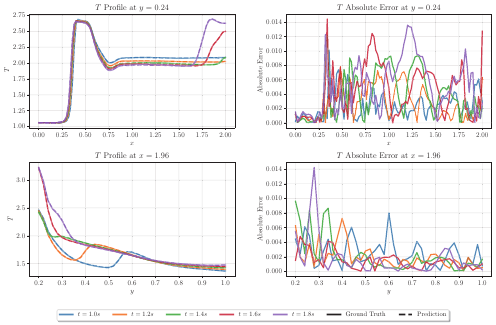}
\caption{$T$ profiles of Forward step in $t=1.0-1.8$ s}
\label{fig.temperature_of_fwdstp}	
\end{figure}

To investigate the detailed features of the flow field, we selected data along two profiles at $y = 0.24$ and $x = 1.96$ for analysis, due to the significant variations occurring near the shock wave. Figure ~\ref{fig.temperature_of_fwdstp} shows the results and errors along these profiles within the time interval $t = 1.0-1.8$ s. It can be observed that the prediction (dashed lines) closely match the ground truth (solid lines). Furthermore, the maximum absolute error not exceeding $0.015$, corresponding to a relative error not exceeding 1\%. This indicates that the model provides accurate and reliable predictions in capturing the interfacial regions near the shock wave.

Table~\ref{table.fwdstp_cost_accuracy} demonstrates the superior efficiency of FNO compared to the fluid solver, with $\mathcal{L}_{\text{r}}(v)$ reaching 0.82\% while $\mathcal{L}_{\text{r}}$ for other physical components remains below 0.55\%. This indicates that our model not only significantly outperforms traditional numerical methods in terms of computational efficiency but also maintains errors at an acceptably low level.

\begin{table}[!ht]
    \centering
    \caption{Computation cost and accuracy comparison of Forward step, TT is abbreviation of the training time of neural networks, IT is the abbreviation of the inferring time of trained models, and CT is the abbreviation of the computational time of Fluid solver. $\mathcal{L}_{\text{r}}$ and residuals are taken as the form: ($p,T,u,v$)}\label{table.fwdstp_cost_accuracy}
    \begin{tabular}{lcccc}
    \toprule
        \textbf{Models} & \textbf{Parameters (Grid Points)} & \textbf{TT} & \textbf{IT (CT)} & \textbf{$\mathcal{L}_{\text{r}}$ (Residuals) (\%)} \\
    \midrule
        Fluid & 4451 & - & 12s & (0.041, 0.042, 0.034, 0.044) \\
        FNO & 8729492 & 48.6s & 9.4ms & (0.55, 0.34, 0.23, 0.82) \\
    \bottomrule
    \end{tabular}
\end{table}
Through meticulous comparison, we have confirmed that the model achieves accuracy comparable to traditional numerical methods while demonstrating superior computational speed.

\subsection{Cylinder Flow}
\label{subsec.cylinder}
\begin{figure}[!ht]
\centering
\includegraphics[width=0.5\textwidth]{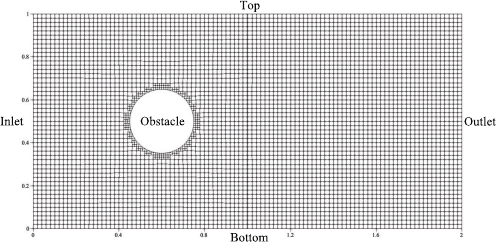}
\caption{Computational grid and boundary labels of Cylinder flow}
\label{fig.cylinder}	
\end{figure}

We conducted an in-depth investigation of the flow field characteristics with a cylindrical obstacle, as illustrated in Fig.~\ref{fig.cylinder}. A cylinder with a radius of 0.15 m is positioned at the center (0.6, 0.5) serving as the obstacle, while the remaining configurations are similar to those described in Section~\ref{subsec.fwdstp}.

The basic computational domain remains discretized with 100 cells in the $x$-direction and 50 cells in the $y$-direction. However, unlike the previous configuration, the cylindrical obstacle causes the surrounding mesh to become non-orthogonal. The initial conditions (internal field) and boundary conditions are detailed in Table~\ref{table.cylinderconfiguration}. Both the internal field and inlet are initialized with a velocity component $u = 1.5-3.0$ m/s, $v = 0$ m/s, pressure $p = 0.7-1.6$ Pa, and temperature $T = 1-1.6$ K, while the boundary conditions remain identical to those described in Section~\ref{subsec.fwdstp}.

\begin{table}[!ht]
    \centering
    \small
    \caption{Boundary conditions and initial field configuration of Cylinder flow}
    \begin{tabular}{lcccc}
    \toprule
        Boundary & $p$ (Pa) & $T$ (K) & $u$ (m/s) & $v$ (m/s) \\ 
    \midrule
        Internal Field/Inlet & 0.7--1.6 & 1.0--1.6 & 1.5--3.0 & 0 \\ 
        Outlet & WT & ZG & ZG & ZG \\ 
        Obstacle & ZG & ZG & slip & slip \\ 
        Top/Bottom & SP & SP & SP & SP \\
    \bottomrule
    \end{tabular}
    \label{table.cylinderconfiguration}
\end{table}

The flow field data is generated using the fluid solver in \textit{OpenFOAM} v11. The computational time step is set to $1 \times 10^{-4}$~s, with the simulation running from 0 to 1~s. Numerical results are sampled every 0.1~s, thereby generating a dataset composed of approximate solutions to the Euler equations. The dataset is partitioned into a training set (4050 samples), a validation set (450 samples), and a test set (600 samples).

\begin{figure}[!ht]
\centering
\includegraphics[width=0.5\textwidth]{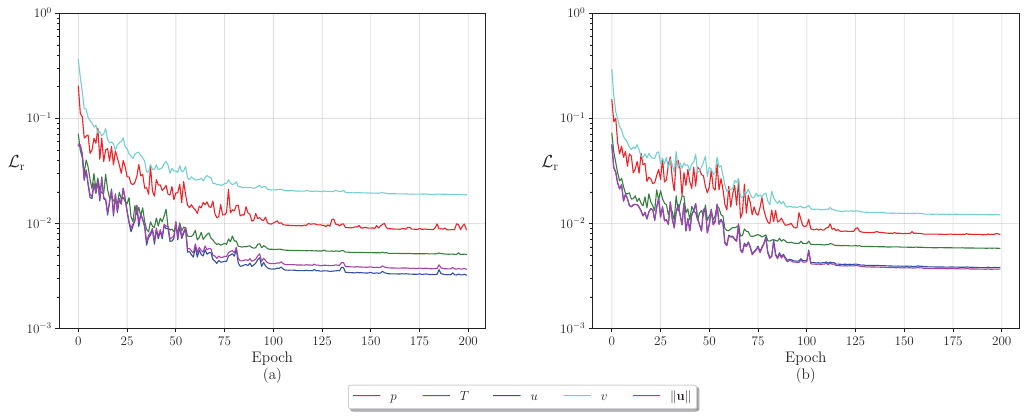}
\caption{Errors of Cylinder flow}
\label{fig.cylinder_loss}
\end{figure}

As shown in Fig.~\ref{fig.cylinder_loss}, subfigure (a) corresponds to the vanilla hyperparameter setting ($\lambda_p = \lambda_T = \lambda_u = \lambda_v = 1$, $\alpha = 1$, and $\beta = 0$), while subfigure (b) adopts the refined configuration ($\lambda_p=1.5$, $\lambda_T=1.0$, $\lambda_u=1.0$, $\lambda_v=2.0$, $\alpha=0.3$, $\beta=0.7$). The results clearly demonstrate that in (a), the errors of $p$ and $v$ remain at elevated levels, with the error of $v$ particularly exceeding 1\% by a considerable margin. In contrast, the optimized hyperparameters in (b) effectively balance the convergence across all physical quantities and significantly reduce the discrepancy in $v$, achieving a more balanced convergence behavior. Additionally, in (b), the gap between the errors of $u$ and $\|\mathbf{u}\|$ is smaller compared to (a), indicating improved convergence of the overall velocity field under the composite loss function.

\begin{figure}[!ht]
\centering
\includegraphics[width=0.5\textwidth]{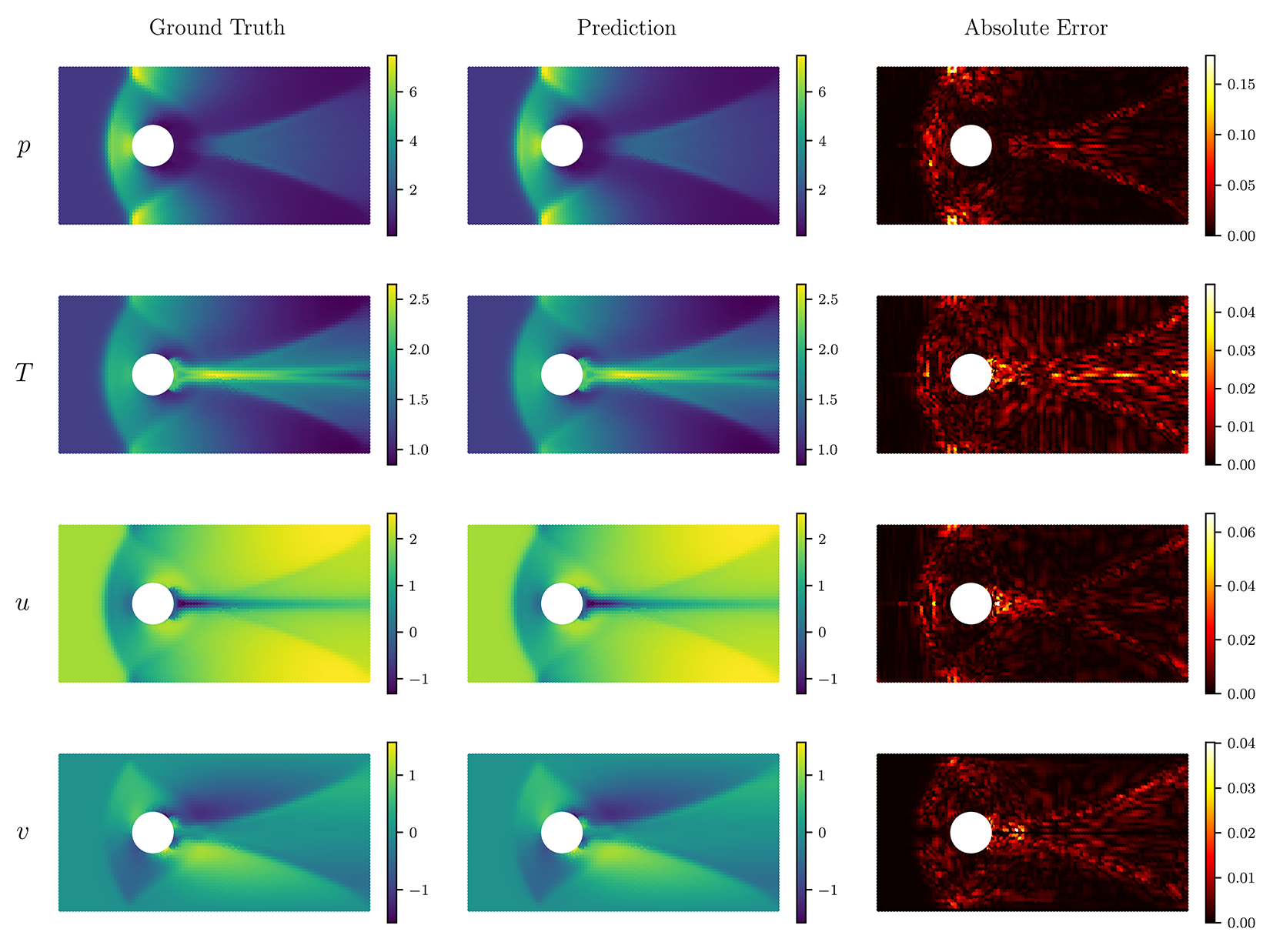}
\caption{Physical field and absolute error of Cylinder flow at $t = 0.9$ s}
\label{fig.cylinder_t9}
\end{figure}

Figure~\ref{fig.cylinder_t9} displays the ground truth, prediction, and absolute errors of the entire physical fields ($p$, $T$, $u$, $v$) at time $t = 0.9$ s. It can be observed from the figure that the predictions are in close agreement with the ground truth, with errors primarily concentrated at the shock locations. The maximum absolute error does not exceed $\left(0.2, 0.05, 0.07, 0.04\right)$.

\begin{figure}[!ht]
\centering
\includegraphics[width=0.5\textwidth]{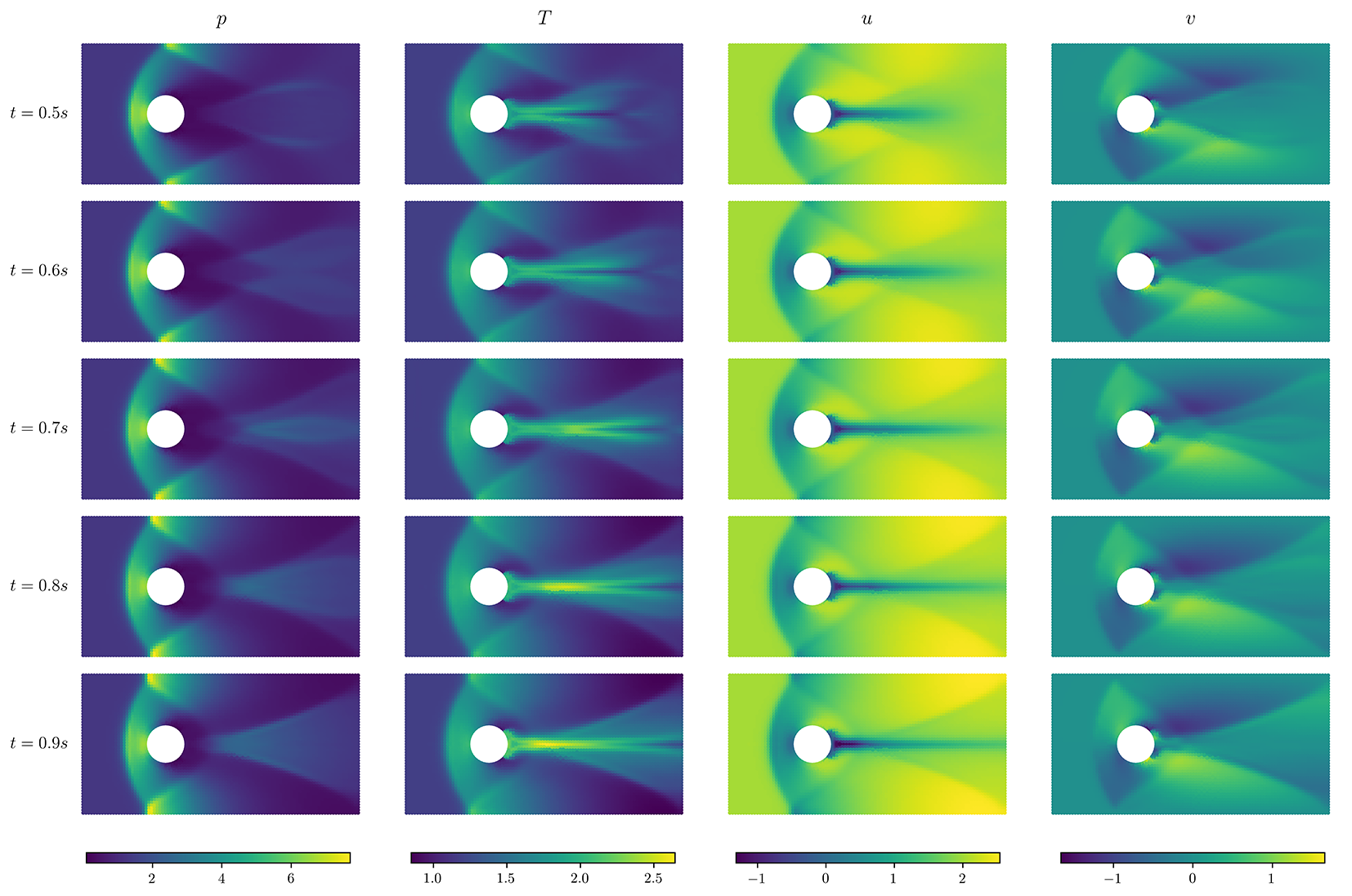}
\caption{Multi-physics field predictions across $t=0.5-0.9$ s of Cylinder flow}
\label{fig.cylindertimeline}	
\end{figure}

We selected a sample from the test set and visualized the predicted flow fields for the four physical quantities—pressure ($p$), temperature ($T$), and velocity components ($u$, $v$)—over the time interval from $t = 0.5 s$ to $t = 0.9 s$. As illustrated in Figure~\ref{fig.cylindertimeline}, the axisymmetry of the flow field causes the shock waves to reflect and converge on the right side of the cylindrical obstacle, leading to intense variations in this region. This visualization allows us to observe the dynamic evolution of each physical quantity over time.

\begin{figure}[!ht]
\centering
\includegraphics[width=0.5\textwidth]{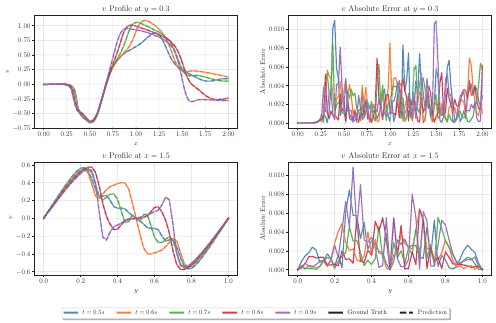}
\caption{$v$  Profiles of Cylinder flow in $t=0.5-0.9$ s}
\label{fig.cylinder_v_profiles}	
\end{figure}

We now focus on the performance of the velocity component $v$ in the flow field. As illustrated in Fig.~\ref{fig.cylinder_v_profiles}, two profiles at $y = 0.3$ and $x = 1.5$ were selected in $t = 0.5 - 0.9$ s due to the significant flow variations observed in these regions. The error distribution analysis reveals that the predicted values closely align with the actual values, with the maximum absolute error not exceeding $0.012$ even in areas of intense flow activity.

As demonstrated in Table~\ref{table.cylinder_cost_accuracy}, the model significantly outperforms traditional numerical methods in terms of computational speed. This enhanced efficiency has substantial implications for time-sensitive applications and large-scale simulations, positioning the model as a promising tool for accelerating compressible flow field analysis. Furthermore, it is noteworthy that the model not only achieves remarkable efficiency but also maintains considerable accuracy in its predictions.

\begin{table}[!ht]
    \centering
    \caption{Computation cost and accuracy comparison of Cylinder flow}

    \begin{tabular}{lcccc}
    \toprule
        \textbf{Models} & \textbf{Parameters (Grid Points)} & \textbf{TT} & \textbf{IT (CT)} & \textbf{$\mathcal{L}_{\text{r}}$ (Residuals) (\%)} \\
    \midrule
        Fluid & 5226 & - & 72.3s & (0.011, 0.014, 0.0096, 0.017) \\
        FNO & 8729492 & 61.3s & 9.8ms & (0.78, 0.57, 0.36, 1.17) \\
    \bottomrule
    \end{tabular}%
	
    \label{table.cylinder_cost_accuracy}
\end{table}

Our numerical simulations on non-orthogonal grids further validate the model's potential, while comprehensive comparisons of flow field characteristics and computational costs collectively underscore its accuracy, adaptability, and innovativeness.

\subsection{Airfoil Flow}
\label{subsec.af}

\begin{figure}[!ht]
\centering
\includegraphics[width=0.5\textwidth]{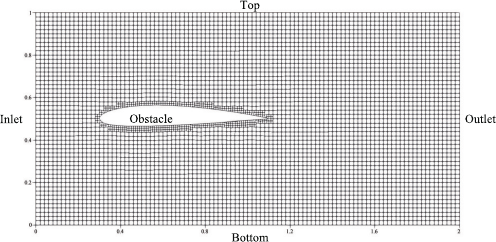}
\caption{Computational grid and boundary labels of Airfoil flow}
\label{fig.airfoil}	
\end{figure}

The flow field described in Section~\ref{subsec.cylinder} is axisymmetric about the horizontal axis. As shown in Fig.~\ref{fig.airfoil}, we now introduce an airfoil (NACA 2412) profile as the obstacle, which generates an asymmetric flow field. The remaining configuration parameters are consistent with those in Section~\ref{subsec.cylinder}, with the initial and boundary conditions detailed in Table~\ref{table.airfoilconfiguration}.

\begin{table}[!ht]
    \centering
    \small
    \caption{Boundary conditions and initial field configuration of Airfoil flow}
    \begin{tabular}{lcccc}
    \toprule
        Boundary & $p$ (Pa) & $T$ (K) & $u$ (m/s) & $v$ (m/s) \\ 
    \midrule
        Internal Field/Inlet & 0.7--1.6 & 1.0--1.6 & 1.5--3.0 & 0 \\ 
        Outlet & WT & ZG & ZG & ZG \\ 
        Obstacle & ZG & ZG & slip & slip \\ 
        Top/Bottom & SP & SP & SP & SP \\
    \bottomrule
    \end{tabular}
    \label{table.airfoilconfiguration}
\end{table}

The flow field data is generated using the fluid solver in \textit{OpenFOAM} v11. The computational time step is set to $1 \times 10^{-4}$~s, with the simulation running from 0 to 1~s. Numerical results are sampled every 0.1~s, thereby generating a dataset composed of approximate solutions to the Euler equations. The dataset is partitioned into a training set (4050 samples), a validation set (450 samples), and a test set (600 samples).

\begin{figure}[!ht]
\centering
\includegraphics[width=0.5\textwidth]{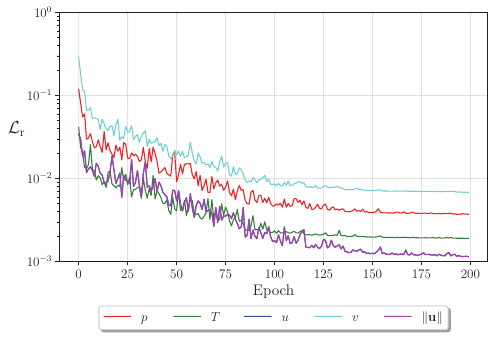}
\caption{Errors of Airfoil flow}
\label{fig.airfoil_loss}
\end{figure}

In Fig.~\ref{fig.airfoil_loss}, with the hyperparameters set as $\lambda_p=1.5$, $\lambda_T=1.0$, $\lambda_u=1.0$, $\lambda_v=2.0$, $\alpha=0.3$, and $\beta=0.7$, it can be observed that the errors of all physical quantities converge to sufficiently low levels. Specifically, the errors for $u$ and $\|\mathbf{u}\|$ are the lowest, nearly approaching 0.001.

\begin{figure}[!ht]
\centering
\includegraphics[width=0.5\textwidth]{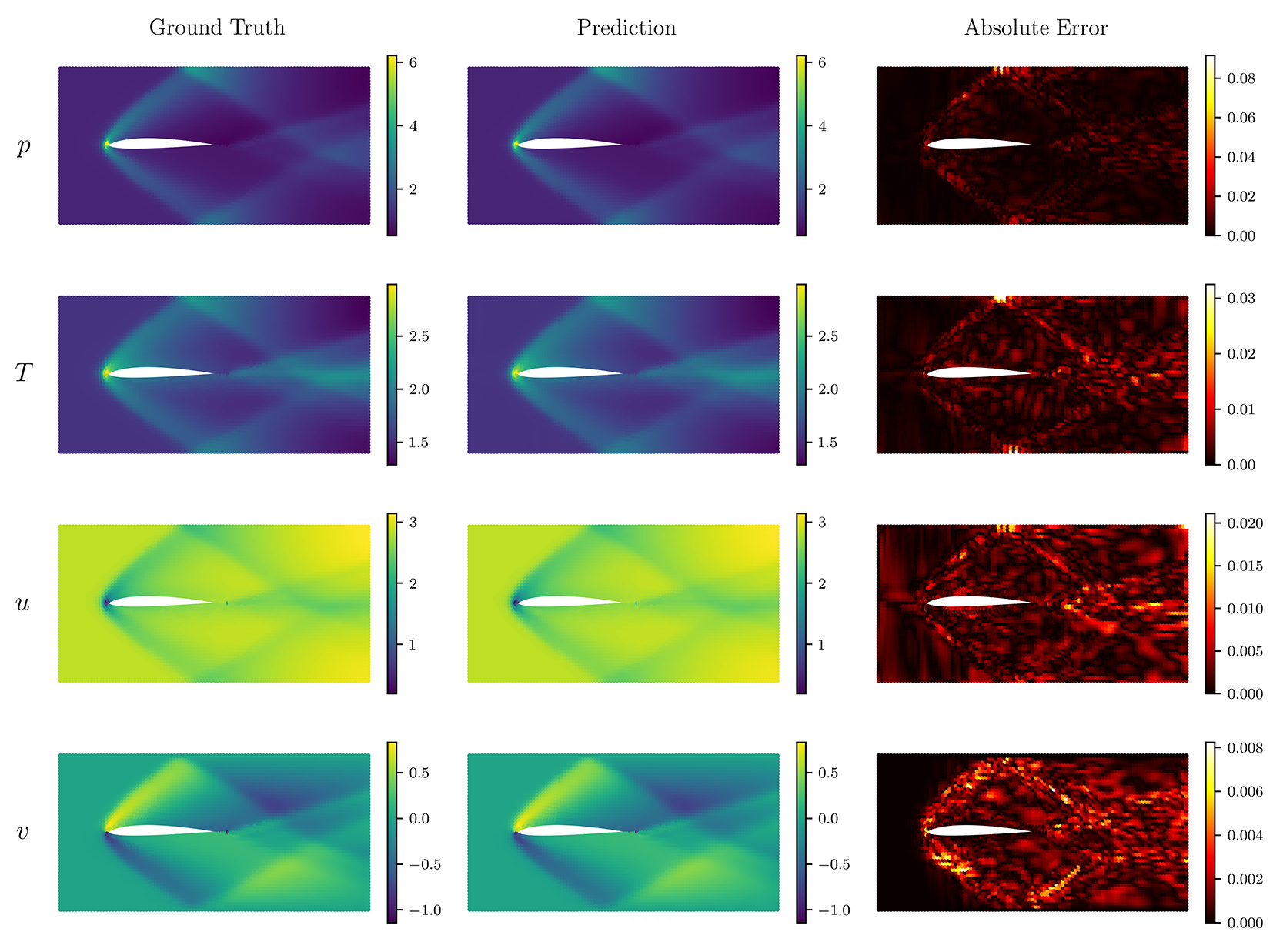}
\caption{Physical field and absolute error of Airfoil flow at $t = 0.9$ s}
\label{fig.airfoil_t9}
\end{figure}

Figure~\ref{fig.airfoil_t9} displays the ground truth, prediction, and absolute errors of the entire physical fields ($p$, $T$, $u$, $v$) at time $t = 0.9$ s. It can be observed from the figure that the predictions closely match the ground truth. The maximum absolute error does not exceed $\left(0.1, 0.04, 0.025, 0.01\right)$.

\begin{figure}[!ht]
\centering
\includegraphics[width=0.5\textwidth]{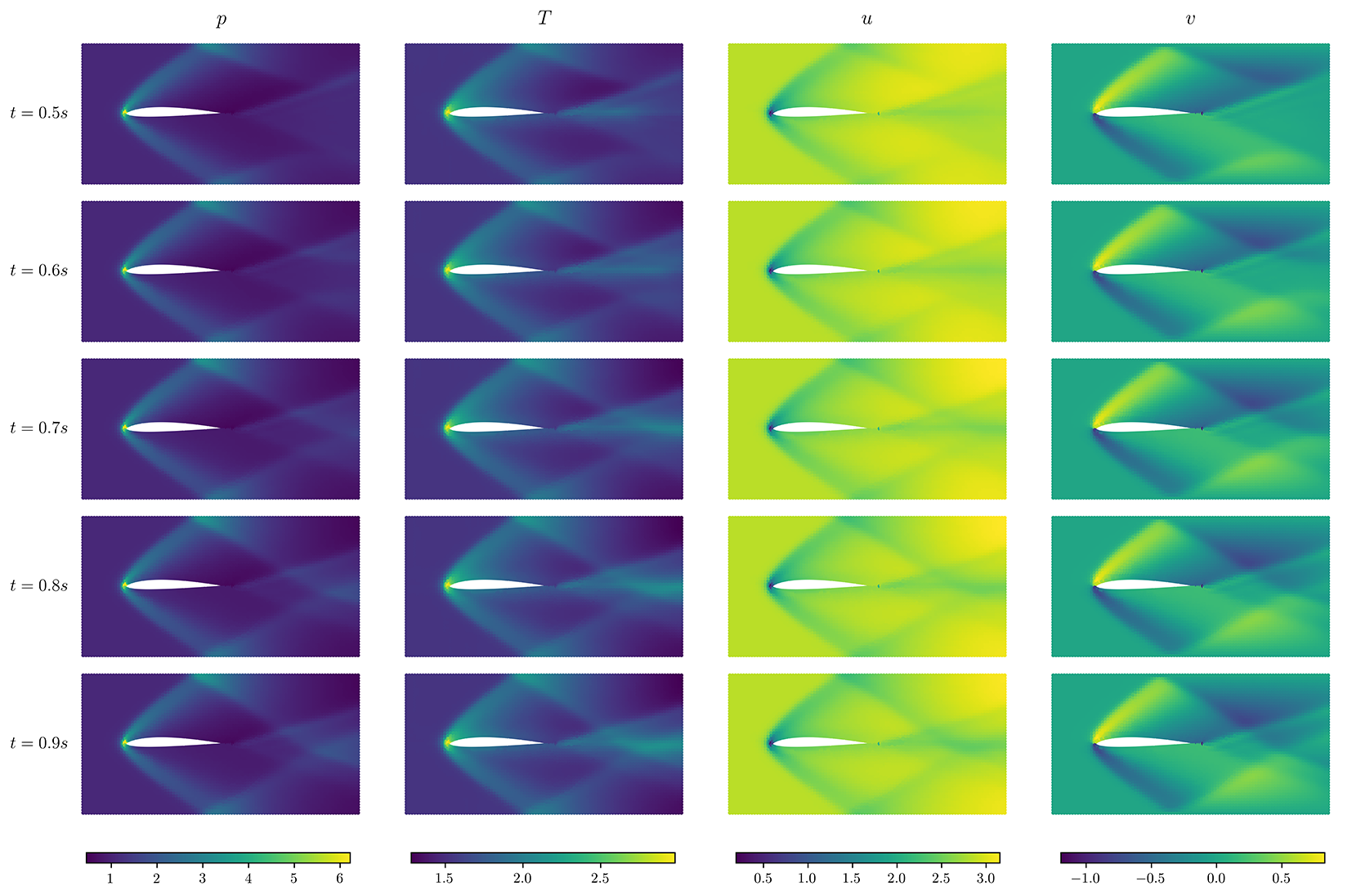}
\caption{Multi-physics field predictions across $t=0.5-0.9$ s of Airfoil flow}
\label{fig.airfoiltimeline}	
\end{figure}

Fig.~\ref{fig.airfoiltimeline} illustrates the temporal evolution of the predicted flow field physical quantities from $t = 0.5$ s to $t = 0.9$ s. The visualization clearly reveals multiple shock wave interactions occurring in the wake region of the flow field.

\begin{figure}[!ht]
\centering
\includegraphics[width=0.5\textwidth]{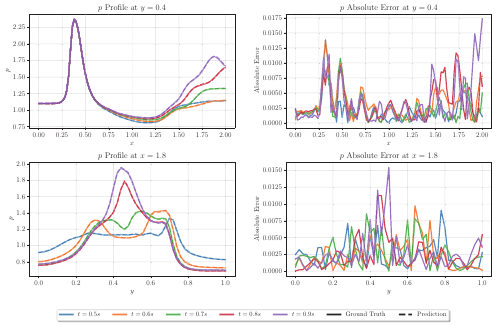}
\caption{$p$ profiles of Airfoil flow in $t=0.5-0.9$ s}
\label{fig.airfoilpressure}	
\end{figure}

Figure~\ref{fig.airfoilpressure} displays the temporal evolution of pressure $p$ along the cross-sectional lines $y = 0.4$ and $x = 1.8$ over the interval $t = 0.5$ s to $t = 0.9$ s. It can be observed that the prediction closely match the ground truth. Furthermore, the absolute error plot on the right reveals a maximum error of less than 1.2\%, underscoring the model's proficiency in capturing flow field dynamics on asymmetric non-orthogonal meshes and affirming its precision and reliability.

As demonstrated in Table~\ref{table.airfoil_cost_accuracy}, the model significantly outperforms traditional numerical methods in terms of computational speed. This enhanced efficiency has substantial implications for time-sensitive applications and large-scale simulations, positioning the model as a promising tool for accelerating compressible flow field analysis. Furthermore, it is noteworthy that the model not only achieves remarkable efficiency but also maintains considerable accuracy in its predictions.

\begin{table}[!ht]
    \centering
    \caption{Computation cost and accuracy comparison of Airfoil flow}
    
    \begin{tabular}{lcccc}
    \toprule
        \textbf{Models} & \textbf{Parameters (Grid Points)} & \textbf{TT} & \textbf{IT (CT)} & \textbf{$\mathcal{L}_{\text{r}}$ (Residuals) (\%)} \\
    \midrule
        Fluid & 5462 & - & 76.6s & (0.023, 0.023, 0.014, 0.026) \\
        FNO & 8729492 & 65.4s & 10.2ms & (0.37, 0.18, 0.12, 0.72) \\
    \bottomrule
    \end{tabular}%
	
    \label{table.airfoil_cost_accuracy}
\end{table}

Our numerical simulations on asymmetric non-orthogonal grids further validate the model's potential. This capability is thoroughly demonstrated through systematic evaluation of flow field behavior and computational performance, collectively affirming the model's precision and flexibility.

\section{Conclusion}
\label{sec.con}

The method proposed in this paper addresses the challenging problem of predicting high-speed compressible fluid flow fields. Notably, all flow fields predicted in this work involve irregular geometries, including both orthogonal and non-orthogonal meshes. By constructing diverse flow field datasets, the model demonstrates robust stability and generalization capability across varying input conditions, achieves remarkable computational efficiency, and delivers competitive accuracy in flow field predictions. These capabilities establish it as an efficient surrogate model, providing a viable path for rapid design evaluation and parameter studies in fields such as aerodynamic shape optimization and propulsion system analysis.

Building upon these demonstrated strengths, several promising research directions emerge for further exploration:

\begin{enumerate}
\item \textbf{Extension to Viscous Flows}: The current work focuses on inviscid flow governed by the Euler equations. Future research should consider extending the methodology to viscous flows described by the Navier-Stokes equations. However, this extension presents significant challenges, including increased complexity in data generation and substantially higher computational demands for numerical simulations.

\item \textbf{High-Dimensional Applications}: In theory, the proposed model can be extended to solve higher-dimensional problems. Nevertheless, such applications are currently constrained by the limited availability of computational resources required for large-scale numerical simulations in elevated dimensions.

\item \textbf{Enhanced Loss Function with Physical Constraints}: The current loss function primarily focuses on the physical quantities themselves. Due to data processing methodologies, calculating derivatives for incorporating physical constraints remains challenging. Future work could develop approaches to overcome this limitation, potentially enabling the integration of governing equations as additional components in the loss function, thereby enhancing physical consistency.
\end{enumerate}

\backmatter

\section*{Declarations}

\bmhead{Funding}

This work was not supported by any funding agency.

\bmhead{Competing interests}

The authors declare no conflict of interest.

\bmhead{Author contributions}

Yifan Nie: Conceptualization, Formal analysis, Methodology, Software, Writing - original draft. Qiaoxin Li: Formal analysis, Methodology, Supervision, Writing - review \& editing.

\bibliography{sn-bibliography}

\end{document}